\documentclass[a4paper]{jpconf}
\usepackage{graphicx}
\usepackage{amssymb,amsmath}

\begin{document}
\title{Perturbative theory of the non-equilibrium singlet-triplet transition}

\author{B. Horv\'ath$^1$, B. Lazarovits$^{2,3}$, G. Zar\'and$^1$}

\address{$^1$Theoretical Physics Department, Institute of Physics, Budapest University of 
Technology and Economics, Budafoki \'ut 8, H-1521 Hungary\\
$^2$Research Institute for Solid State Physics and Optics of the Hungarian Academy 
of Sciences,
Konkoly-Thege M. \'ut 29-33., H-1121 Budapest, Hungary
\\
$^3$Department of Physics, Rutgers University, Piscataway, New Jersey 08854, USA}

%\begin{abstract}
%Multi-level quantum dots have been the subject of intense theoretical and experimental investigation for the rich physics they display: exotic correlated states such as the $SU(4)$ Kondo state or the orbital Kondo effect as well as quantum phase transitions (cross-overs) as the singlet-triplet transition have been observed in quantum dots having almost degenerate orbitals. 

%\section{Abstract}
\begin{abstract}
We study equilibrium and non-equilibrium properties 
of a two-level quantum dot close to the singlet-triplet transition. 
We treat the on-site Coulomb interaction and Hund's rule coupling 
perturbatively within the Keldysh formalism.
We compute the spectral functions and the differential 
conductance of the dot. 
For moderate interactions our perturbative approach captures the Kondo effect 
and many of the experimentally observed properties. 
\end{abstract}
%\maketitle

\section{Introduction}

Handling strongly interacting multilevel 
 systems under non-equilibrium conditions
is of crucial importance for understanding transport properties of 
molecules and  correlated  mesoscopic structures. Although  the theoretical description of these %strongly correlated 
systems  is rather satisfactory in 
equilibrium~\cite{hofstetter1,yeyati2,lichtenstein,zarand,izumida,ali1}, 
the success of non-equilibrium methods is rather 
limited~\cite{paaske}: In fact, most of the available 
 methods are unable to capture the Kondo physics or are rather 
uncontrolled.   Perturbation theory in this regard is of particular importance: 
Although it breaks down for strong interactions, 
for moderate interaction strengths it is able to capture 
the formation of the Kondo resonance and the Hubbard peaks \cite{zlatic}, 
scales very well with the number of orbitals, and 
it is therefore a promising candidate to combine with {\em ab initio} 
calculations. 

Motivated by experiments on  lateral quantum dots~\cite{wiel} and
carbon nanotubes~\cite{jarillo},
here we focus on a particular parameter range of multilevel quantum
dots, and study  the so-called  singlet-triplet transition, i.e., 
the  transition from a singlet state of the dot to a Kondo-screened 
triplet state,  driven by the presence of Hund's rule coupling. 
We show that  the so-called interpolative perturbation 
theory (IPT)\cite{yeyati2,yeyati1,kajueter,ali2} can be
extended to include Hund's rule coupling, and it captures  such basic
features of this transition  as the splitting of the 
Kondo resonance~\cite{wiel}. A detailed analysis of the 
transition shall be published elsewhere~\cite{future}.

\section{Model and theoretical framework}

We use the following two level %Anderson 
Hamiltonian to investigate the transition, 
$H= H_{0}+H_{\rm int}$, 
\begin{eqnarray}
%   H & = & H_{0}+H_{\rm int}\;\label{ham},\\
   H_{0} & = & \sum_{\xi,\alpha,\sigma}\xi_{\alpha}c_{\xi\alpha\sigma}^{\dagger}c_{\xi\alpha\sigma}
         +
         \sum_{i,\sigma}\varepsilon_{i}d_{i\sigma}^{\dagger}d_{i\sigma} 
         + \sum_{\alpha,i,\xi,\sigma}t_{\alpha
           i}(c_{\xi\alpha\sigma}^{\dagger}d_{i\sigma}+h.c.)\;,
\label{ham}
%\nonumber
\\
   H_{\rm int} & = & {U}/{2}\;{\textstyle\sum_{(i,\sigma)\neq(j,\sigma')}}n_{i\sigma}n_{j\sigma'}-J\vec{S}^{2}\;.\nonumber
\end{eqnarray}
Here $c_{\xi\alpha\sigma}^{\dagger}$ creates a conduction electron in the left or right lead with energy 
$\xi_{\alpha}=\xi+\mu_{\alpha}$ ($\mu_{\alpha}=eV_{\alpha}$ is the bias applied on lead $\alpha\in(L,R)$) 
and spin $\sigma$ and $d_{i\sigma}^{\dagger}$ is the creation operator of an electron of spin $\sigma$, 
and energy $\varepsilon_{\pm}$ on the dot level $i\in(+,-)$. 
The $t_{\alpha i}$ denote the tunneling matrix elements between lead $\alpha$ and dot level $i$. 
Throughout this paper we focus on a completely symmetrical lateral quantum dot, and assume that one of the dot 
levels is even ($+$), while the other is odd under reflection ($-$). 
As a consequence, the tunneling matrix elements have a simple structure: $t_{L,+}=t_{R,+}=v_{+}/\sqrt{2}$ and
 $t_{L,-}=-t_{R,-}=v_{-}/\sqrt{2}$ \cite{zarand,izumida}. 
The width of the dot-levels is given by $\Gamma_{\pm}=2\pi\rho_{0}|v_{\pm}|^{2}$, with $\rho_{0}$ the density 
of states of the electrons in the leads. The coupling $U$ denotes the on-site Coulomb interaction, and accounts 
for the charging energy of the dot, while $J$ stands for the Hund's rule coupling which favors a ferromagnetic 
alignment of the total spin of the dot, 
$\vec{S}=\frac{1}{2}\sum_{i\sigma\sigma'}d_{i\sigma}^{\dagger}\vec{\sigma}d_{i\sigma}$.

The simple Hamiltonian (\ref{ham}) describes a variety of physical phenomena, and captures, e. g. the 
underscreened \cite{roch} and the $SU(4)$ Kondo states \cite{jarillo,anders,zarand1,tarucha}. 
Here, however, we shall focus only to the regime with $\langle n_{+}+n_{-}\rangle\approx2$, and the vicinity of 
the singlet-triplet transition induced by the competition of the Hund's rule coupling and the separation of the two levels, 
$\Delta\equiv\varepsilon_{+}-\varepsilon_{-}$.

We shall treat (\ref{ham}) by applying perturbation theory in $U$ and
$J$. However, before doing so, 
we separate the Hartree contribution by introducing counterterms, 
 $H_{0}=\tilde{H}_{0}+H_{\rm count}$, 
\begin{eqnarray}
%   H_{0} & = & \tilde{H}_{0}+H_{\rm count}\;,\\
\label{ham0}
   \tilde{H}_{0} & = & \sum_{\xi,\alpha,\sigma}\xi_{\alpha}c_{k\alpha\sigma}^{\dagger}c_{k\alpha\sigma}
         + \sum_{i,\sigma}\tilde{\varepsilon}_{i\sigma}d_{i\sigma}^{\dagger}d_{i\sigma}
         + \sum_{\alpha,i,\xi,\sigma}t_{\alpha
           i}(c_{\xi\alpha\sigma}^{\dagger}d_{i\sigma}+h.c.)\;,
%\\
% H_{\rm count} & = & \sum_{i,\sigma}(\varepsilon_{i}-\tilde{\varepsilon}_{i\sigma})d_{i\sigma}^{\dagger}d_{i\sigma}\;.\nonumber
\end{eqnarray}
We then use $\tilde{H}_{0}$ to obtain the unperturbed Keldysh Green's functions,
 $g_{ii'\sigma}^{\kappa\kappa'}$, with $i$ and $i'$ dot-level labels and $\kappa$ and $\kappa'=\pm$ 
the usual Keldysh labels, and treat 
$H_{\rm count}= \sum_{i,\sigma}(\varepsilon_{i}-\tilde{\varepsilon}_{i\sigma})d_{i\sigma}^{\dagger}d_{i\sigma}$ as a perturbation.

\begin{figure}[t]
%\begin{center}
\hskip3cm\includegraphics[width=260pt]{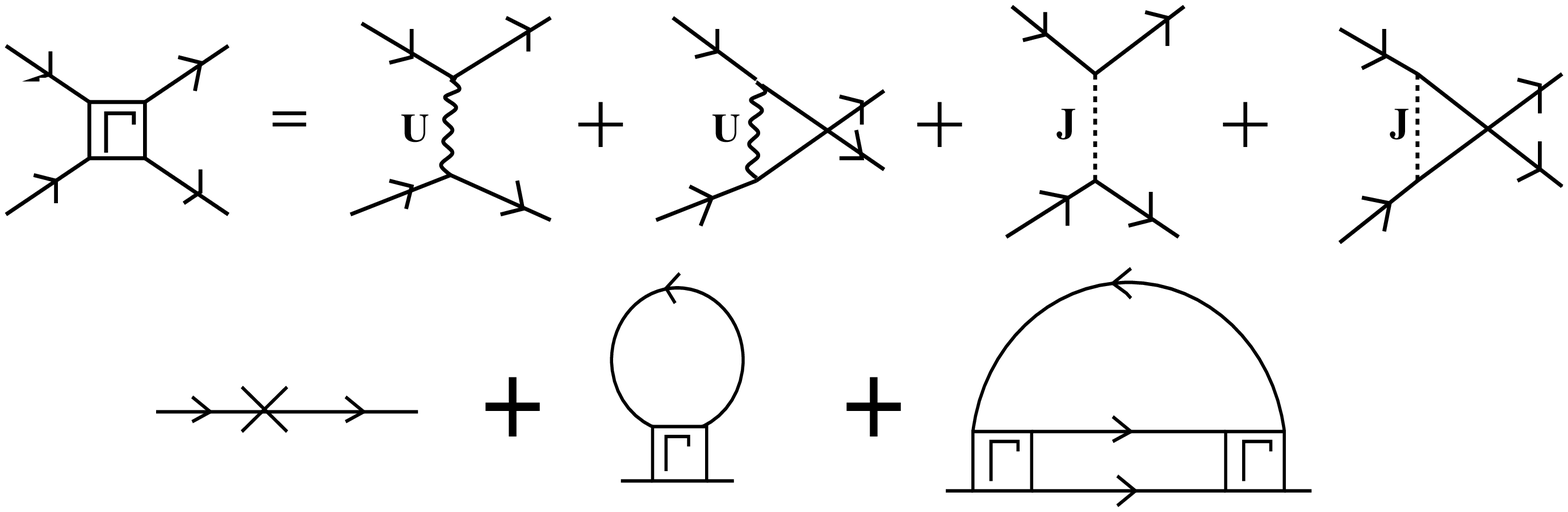}
%\end{center}
%\includegraphics[width=320pt]{diag.eps}
\caption{The symmetrized vertex and the first and second order
  self-energy diagrams}\label{vertexfig}
\end{figure}

We then need to do second order perturbation theory in $H_{\rm int}$. To treat both the Coulomb interaction
 and the Hund's rule coupling on equal footing, we merge them into a single interaction vertex, 
$\Gamma_{i\sigma\;n\tilde\sigma}^{j\sigma'\;m\tilde\sigma'}$, as diagramatically shown in Fig. \ref{vertexfig}.
Then the second order self-energy can be expressed in time domain as follows,
\begin{equation}
   {\Sigma^{(2)}}_{\;ii'\sigma}^{\kappa\kappa'}(t) = \sum_{\substack{j,j',m,m',n,n'\\\sigma', \sigma'',\sigma'''}}\Gamma_{i\sigma\;n\sigma'''}^{j\sigma'\;m\sigma''}\Gamma_{j'\sigma'\;m\sigma''}^{i'\sigma \;n'\sigma'''}\;g_{jj'\sigma'}^{\kappa\kappa'}(t)\;g_{mm'\sigma''}^{\kappa\kappa'}(t)\;g_{n'n\sigma'''}^{\kappa'\kappa}(-t)\label{2ndself}\;.
\end{equation}
Up to second order, the frequency dependent Green's functions can be
obtained from the following Dyson's equation
\begin{equation}
   \left[ G_{ii'\sigma}^{\kappa\kappa'}(\omega)\right]^{-1} = \left[ g_{ii'\sigma}^{\kappa\kappa'}(\omega)\right]^{-1}-{\Sigma^{(1)}}_{ii'\sigma}^{\kappa\kappa'}(\omega)-{\Sigma^{(2)}}_{ii'\sigma}^{\kappa\kappa'}(\omega)\label{dys}\;,
\end{equation}
where ${\Sigma^{(1)}}_{ii'\sigma}^{\kappa\kappa'}(\omega)$ contains the self-energy part coming from the counterterm and the first order Hartree contribution, also shown in Fig. \ref{vertexfig}.

Eqs. (\ref{2ndself}) and (\ref{dys}) give a complete perturbative
description of the quantum dot, however, they depend parametrically on
the so far unspecified levels,
$\tilde{\varepsilon}_{\pm\sigma}$. These are determined 
selfconsistently from Eqs.~ (\ref{2ndself}) and (\ref{dys})
by requiring that $g$ and $G$ give the same occupation numbers \cite{yeyati2}
\begin{equation} 
   n_{i\sigma}^{(0)}\left[ \tilde{\varepsilon}_{i\sigma}\right] = \frac{1}{2\pi i}\int_{-\infty}^{\infty}g_{ii\sigma}^{12}(\omega)d\omega \equiv \frac{1}{2\pi i}\int_{-\infty}^{\infty}G_{ii\sigma}^{12}(\omega)d\omega = n_{i\sigma}^{(2)}\left[ \tilde{\varepsilon}_{i\sigma}\right]\label{occup}\;.
\end{equation}
%Equations (\ref{2ndself}), (\ref{dys}) and (\ref{occup}) are then solved iteratively.

\begin{figure}[b]
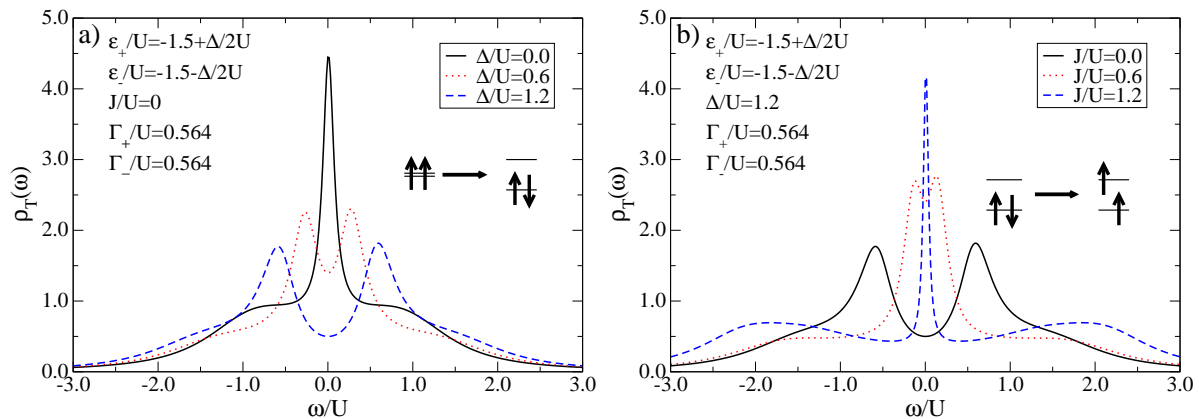

\includegraphics[width=220pt,clip]{dvar.eps}
\includegraphics[width=220pt,clip]{jvar.eps}
\caption{Total spectral function, $\rho_{T}(\omega)\equiv\sum_{i,\sigma}\rho_{i\sigma}(\omega)$. Evolution of (a) spectral functions as a function of level splitting, $\Delta$, (b) spectral functions as a function of Hund's rule coupling, $J$.}\label{dvarfig}
\end{figure}

%\section{Results}
{\parindent=0pt {\em  Equilibrium results}: For simplicity, here we focus on the electron-hole symmetric case, where $\varepsilon_{\pm}=-1.5U\pm\Delta/2$. First we consider the equilibrium spectral functions of the two levels:
\begin{equation}
   \rho_{i\sigma}(\omega)=-\frac{1}{\pi}\Im m\;G_{\;ii\sigma}^{R}(\omega)=-\frac{1}{2\pi i}\left( G_{\;ii\sigma}^{>}(\omega)-G_{\;ii\sigma}^{<}(\omega)\right)\;,
\end{equation}
the sum of which, $\rho_{T}(\omega)\equiv\sum_{i,\sigma}\rho_{i\sigma}(\omega)$, is plotted in Fig. \ref{dvarfig}. Consider first the case $J=0$ (Fig. \ref{dvarfig}a). In the absence of hybridization, the ground state is highly degenerate for $\Delta=0$, and turning on a finite hybridization leads to the appearance of a Rondo resonance. Changing the value of the level splitting ($\Delta/U$) the two electrons are forced to stay on the lower level, $\varepsilon_{-}$, and a dip opens in the spectral function, corresponding to a singlet ground state.}

In Fig. \ref{dvarfig}b we analyze the effect of Hund's rule coupling, $J$. For $J\gtrapprox\Delta$ a triplet ground state is favored, and a Kondo effect develops, where the spin $S=1$ of the dot is screened by the even and odd combinations of the conduction states. A clear signal of the singlet-triplet transition is that the zero bias dip of the spectral function closes and a Kondo peak develops on the triplet side. These features are surprisingly well captured by the simple perturbative calculation presented here.

\begin{figure}
\begin{center}
\includegraphics[width=220pt,clip]{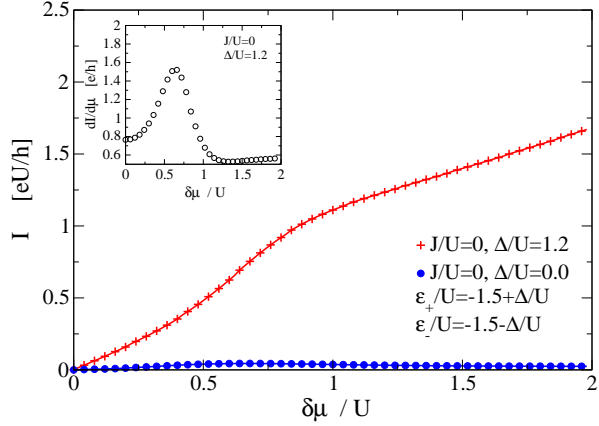}
\caption{Current versus dimensionless bias,
  $\delta\mu/U=eV/U$, for a singlet ground state ($J/U=0$,
  $\Delta/U=1.2$) and for the highly degenerate ground state ($J/U=0$,
  $\Delta/U=0$).
%Linear fits (dashed lines) were obtained from Fermi-liquid
% relations.
 Inset: differential conductance for $J/U=0$, $\Delta/U=1.2$. In all
 cases $\Gamma_{+}=\Gamma_{-}=\Gamma=0.564\;U$.}\label{curfig} 
\end{center}
\end{figure}

{\parindent=0pt {\em Out of equilibrium results:} Equations
(\ref{2ndself}), (\ref{dys}) and (\ref{occup}) also provide a closed
set of equations for the local Green's function in
non-equilibrium. Solving them we can then compute transport properties
of the dot using the Meir--Wingreen formula \cite{meir}. 
%:
%\begin{equation}
%   I = \frac{ie}{2h}\int d\omega\left[\mathrm{Tr}\left\lbrace (f_{L}(\omega)\mathbf{\Gamma}^{L}-f_{R}(\omega)\mathbf{\Gamma}^{R}) (\mathbf{G}^{>}(\omega)-\mathbf{G}^{<}(\omega))\right\rbrace +\mathrm{Tr}\left\lbrace (\mathbf{\Gamma}^{L}-\mathbf{\Gamma}^{R})\mathbf{G}^{<}(\omega)\right\rbrace \right]\;,\label{curexp}
%\end{equation}
%where $f_{\alpha}(\omega)\equiv f(\omega-\mu_{\alpha})$, and the bold
%letters in the expression denote matrices in dot-level indices,
%$\mathbf{\Gamma}_{i\sigma,i'\sigma'}^{\alpha}\equiv\delta_{\sigma\sigma'}2\pi\rho_{0}t_{\alpha
%i}^{*}t_{\alpha i'}$. The trace is carried out over level and spin
%indices. 
In our calculations the effective energies were fixed in equilibrium 
for a given parameter set, and we used these equilibrium values 
for all bias voltages.}

In Fig. \ref{curfig} the current is plotted as a function of bias,
$\delta\mu=\mu_{L}-\mu_{R}=eV$, for two different parameter sets. In
case of $J/U=0$ and $\Delta/U=0$ the current is suppressed and the
linear conductance vanishes. This is due to the destructive
interference between the even and odd channels, which both acquire a
phase shift $\delta_{\pm}=\pi/2$ in this Kondo regime. The other
current plot (red crosses) corresponds to a dot with singlet ground
state. The differential conductance displays a dip at zero bias (see
inset of Fig. \ref{curfig}), which is a clear fingerprint of the
non-equilibrium singlet-triplet transition. 
%As indicated by the dashed lines, the initial slope of the total
%current is directly related to the scattering phase shifts that we
%determined from the equilibrium spectral functions using Fermi liquid
%relations \cite{future,hewson}.

\section{Conclusions}
In summary, we find that 
self-consistent second order 
perturbation theory captures qualitatively
the  transport properties of a two-level quantum dot throughout the
singlet-triplet transition, and the splitting of the Kondo resonance. 

\section{Acknowledgments}

This research has been supported by Grants No. OTKA 
NF061726, F68726, NN76727 and NSF-DMR 0806937.

%\begin{figure}
%\includegraphics[width=300pt]{dvar.eps}
%\caption{Spectral functions beginning from a highly degenerate ground state and reaching a singlet state by changing level splitting ($\Delta$)}\label{dvarfig}
%\end{figure}

\end{document}